\documentclass[12pt,preprint,showpacs,aps]{revtex4}
\usepackage{epsfig}
\usepackage{amsmath}
\usepackage{mathrsfs}

\begin{document}

\title{Quasinormal modes of plane-symmetric
anti-de Sitter black holes: A complete analysis of
the gravitational perturbations}

\author{Alex S. Miranda\footnote{amiranda@mail.ufsm.br}
and Vilson T. Zanchin\footnote{zanchin@ccne.ufsm.br}}
\affiliation{Departamento de F\'{\i}sica, Universidade
Federal de Santa Maria, 97119-900 Santa Maria, RS, Brazil.}

\begin{abstract}

We study in detail the quasinormal modes of linear
gravitational perturbations of plane-symmetric anti-de
Sitter black holes. The wave equations are obtained by
means of the Newman-Penrose formalism and the Chandrasekhar
transformation theory. We show that oscillatory modes decay
exponentially with time such that these black
holes are stable against gravitational perturbations.
Our numerical results show that in the large (small)
black hole regime the frequencies of the ordinary quasinormal
modes are proportional to the horizon radius $r_{+}$ (wave number $k$).
The frequency of the purely damped mode is very close to the
algebraically special frequency in the small horizon limit,
and goes as $ik^{2}/3r_{+}$ in the opposite limit. This
result is confirmed by an analytical method based on the
power series expansion of the frequency in terms of the
horizon radius. The same procedure applied to the
Schwarzschild anti-de Sitter spacetime proves that the
purely damped frequency goes as $i(l-1)(l+2)/3r_{+}$,
where $l$ is the quantum number characterizing the
angular distribution. Finally, we study the limit of
high overtones and find that the frequencies become evenly
spaced in this regime. The spacing of the frequency per unit horizon
radius seems to be a universal quantity, in the sense
that it is independent of the wave number, perturbation
parity and black hole size.

\end{abstract}


\maketitle

\section{Introduction}
\label{introducao}

The dynamical response of black holes to any external
disturbance is dominated, at intermediate times, by
a discrete set of damped oscillations called quasinormal modes
(QNMs). This result has been verified not only at linearized
level, but also in fully numerical simulations for stellar
gravitational collapse \cite{star} and collision of two black
holes \cite{anni}. The possibility of observation in the
forthcoming years of the radiation due to the QNMs opens a new
perspective for the gravitational wave astronomy. From these
observations it will be possible to obtain direct evidence of
black holes' existence, as well as to estimate their parameters
(mass, electric charge, and angular momentum) \cite{kokk}.

In addition to the astrophysical applications, an interpretation
appeared some years ago of the QNMs in terms of the AdS/CFT
conjecture \cite{horo}. According to this conjecture, there is a
correspondence relation (or duality) between string theory in an
anti-de Sitter (AdS) spacetime and a conformal field theory (CFT) on
the boundary of this space \cite{mald,ahar}. It implies that a large
black hole in the AdS spacetime corresponds to an approximately
thermal state in the CFT. Perturbing the black hole is equivalent to
perturbing the thermal state, and the perturbation damping time gives
the time scale for the return to the thermal equilibrium. Therefore, by
computing quasinormal (QN) frequencies in an AdS space it is possible
to obtain the thermalization time scale in the strongly coupled CFT.

The early works studying perturbations of AdS black holes considered the
evolution of a conformally invariant scalar field in such background
spacetimes \cite{cha1,cha2}. After the advent of the AdS/CFT
conjecture, QNMs of spacetimes with a negative cosmological constant,
such as Kottler (also called Schwarzschild-AdS)
\cite{horo,car1,govi,kon1,moss,mus1,giam,jing1},
Reissner-Nordstr\"om-AdS \cite{wan1,kon2,bert1,kon3,jing2} and
Kerr-AdS \cite{car2,car3,bert2}, were computed for different fields,
dimensions, and boundary conditions. Within this program, other kinds
of black hole solutions were also analyzed. For instance, in the
Ba\~nados-Teitelboim-Zanelli solution \cite{bana}, analytical
expressions for the QN frequencies presented exact agreement with the
poles of the retarded correlation function in the dual CFT
\cite{car4,birm}. This result provided an important quantitative test
of the AdS/CFT correspondence. As a complement to the aforementioned
studies, analyses of the time evolution of some fields in
asymptotically AdS backgrounds have been performed
\cite{wan2,zhu,wan3,wan4}. More recently, the conjecture relating the
highly damped QNMs to black hole area quantization has motivated a
search for QN frequencies with large imaginary parts also in
asymptotically AdS spacetimes
\cite{car5,mus2,siop,car6,nata,setare1}.

The studies of perturbations in AdS spacetimes were also extended to
black holes with nonspherical topology. In Ref. \cite{cha2} the
evolution of a conformal scalar field in the family
of topological backgrounds was investigated, which includes the
plane-symmetric black hole as one of its members, while in Ref.
\cite{wan3} the case of a nonminimally coupled scalar field in the
same kind of spacetimes was considered. Lately, the lowest QN frequencies
associated with the scalar, electromagnetic, and gravitational
perturbations of the plane-symmetric black hole background were computed
\cite{car7}. The study of the scalar and electromagnetic fields was
relatively complete. However, important questions were left without
answers in connection with the gravitational perturbations. For
instance, what are the wave equations governing general metric
perturbations? What is the physical interpretation of the modes with
different wave numbers? What are the QN frequencies in the intermediate
and small black hole regimes? What is the relation between these
frequencies and the wave number? Why is the pure imaginary frequency
not proportional to the horizon radius in the large black hole limit
as it happens for ordinary modes? Is there some relation between the
latter and the algebraically special frequency? And what is the behavior
of the QNM spectrum in the regime of highly damped overtones?

The objective of the present work is answering most of the previous
questions as well as other ones that appeared during this work.
Some of the conclusions drawn here are extended to the
Schwarzschild-AdS black hole. The plan of this paper is the
following. In Sec. \ref{solucao}, we write the equations
defining the background spacetime in a
Newman-Penrose (NP) null tetrad frame \cite{newm}. In
Sec. \ref{perturb-NP}, we obtain Schr\"odingerlike equations
governing the axial and polar perturbations. In Sec. \ref{qnm}, we
define the QNM boundary conditions to be used and prove the black
holes stability against perturbations with a non-negative
potential. A brief description of the Horowitz-Hubeny
method (Sec. \ref{metodo}) follows, as well as subsequent
exposition of the QN frequency values for different horizon
radii and wave numbers (Sec. \ref{resultados}).
Then, we study the large black hole limit of
the axial QNMs (Sec. \ref{limite}) and the non-QN character of the
algebraically special frequency (Sec. \ref{algebrico}). Finally, we
conclude by summarizing and commenting the main results
(Sec. \ref{conclusao}).

Throughout this work, we use the geometric system of units in which
the speed of light and the Newton's gravitational constant are set to
unity. The background spacetime, moreover, has a natural lengthscale
given by the AdS radius $R=\sqrt{3/|\Lambda_{c}|}$, where
$\Lambda_{c}$ is the negative cosmological constant. Here we choose
the length unit so that $R=1$ and measure everything in terms of the
AdS radius. In relation to notation, the asterisk ($\ast$) denotes
complex conjugation and the definition of the NP quantities follows
Ref. \cite{chan1}.

\section{The Black Hole Spacetime}
\label{solucao}

With the aim of investigating the evolution of linear gravitational
perturbations in the background of a plane-symmetric AdS black hole
\cite{kram}, we summarize here the basic properties of such a
spacetime. The metric may be written in the form
\begin{equation}
ds^{2}= f(r)dt^{2}-f(r)^{-1}dr^{2}-
r^{2}(d\varphi^{2}+dz^{2}),
\label{toroidal}
\end{equation}
where 
\begin{equation*}
f(r)=r^{2}-4M/r.
\end{equation*}
There is an event horizon at $r =r_{+}=(4M)^{1/3}$, and an essential
singularity at $r=0$. Depending on the ranges of coordinates $\varphi$
and $z$, the metric may represent a cylindrical ($0\leq\varphi <
2\pi$, $-\infty<z< +\infty$), toroidal ($0\leq\varphi,z< 2\pi$)
or planar ($-\infty<\varphi,z< +\infty$) black hole
spacetime. For the torus, $M$ is related to the ADM system mass
\cite{huan}; for the cylinder, to the mass per length unit of $z$ line
\cite{lemo,lemosz}; and for the plane, to the mass per area unit of the
$\varphi-z$ plane \cite{cai}.

According to the Petrov classification, the above metric
\eqref{toroidal} is algebraically type D. It is then convenient to define
the NP quantities in terms of a Kinnersley-like null frame
\cite{kinn}. In this case, the two real vectors of such a frame,
$D$ and $\Delta$, are chosen to lie, respectively, on the ingoing and
outgoing radial null geodesics and are defined by
\begin{gather}
D=l^{\mu}\partial_{\mu}=\frac{1}{f}\left(\partial_{t}+
f\partial_{r}\right),
\label{vec-l}\\
\Delta=n^{\mu}
\partial_{\mu}=\frac{1}{2}(\partial_{t}-
f\partial_{r}),
\label{vec-n}
\end{gather}
which are the double principal null directions of the Weyl tensor.
In order to complete the frame basis we have the vector
\begin{equation}
\delta=m^{\mu}\partial_{\mu}=\frac{1}{r\sqrt{2}}
(\partial_{\varphi}+i\partial_{z}),
\label{vec-m}
\end{equation}
and its complex conjugate $\delta^{\ast}=m^{\ast\mu}\partial_{\mu}$.
Both of these vectors, $m^\mu$ and $m^{\ast\mu}$, are orthogonal to
$l^{\mu}$ and $n^{\mu}$, and satisfy the normalization condition
$m^{\mu}m_{\mu}^{\ast}=-1$. With the chosen tetrad basis, the only
nonvanishing spin coefficients are
\begin{equation}
\rho=-1/r,\qquad\quad\mu=-f/2r,\qquad\quad\gamma=(r^{3}+2M)/2r^{2},
\label{spins}
\end{equation}
and the only nonvanishing Weyl and Ricci scalars are
\begin{equation}
\Psi_{2}=-2M/r^{3},\qquad\quad\Lambda=1/2.
\label{curvs}
\end{equation}

\section{The wave equations}
\label{perturb-NP}

Here the equations for the linear gravitational perturbations are
obtained by using the NP formalism, which allows us to investigate
general properties of such perturbations without any further
simplifying assumption. In this approach, changes in the metric
coefficients are directly related to changes in the tetrad null
vectors. Furthermore, perturbations in the vectors $l^{\mu}$,
$n^{\mu}$ and $m^{\mu}$ lead to first order alterations in Weyl and
Ricci scalars, and also in the spin coefficients. The full system of
linearized perturbation equations is obtained by expanding the complete 
NP equations on the plane-symmetric black hole background,
and retaining only the
first order terms in the perturbation functions.

Generally, the linearized NP equations form a large
system of coupled equations whose solutions are difficult
to be found. However, for Petrov type D vacuum spacetimes,
the problem reduces to solving a pair of decoupled equations
for the gauge and tetrad invariants $\Psi_{0}$ and $\Psi_{4}$
\cite{teuk}. In the present case, these equations assume
the simple forms
\begin{gather}
[(D-4\rho-\rho^{\ast})(\Delta-4\gamma+\mu)-
\delta\delta^{\ast}-3\Psi_{2}]\Psi_{0}=0,
\label{fund1}\\
[(\Delta+3\gamma-\gamma^{\ast}+4\mu+\mu^{\ast})(D-\rho)-
\delta^{\ast}\delta-3\Psi_{2}]\Psi_{4}=0.
\label{fund2}
\end{gather}
The foregoing equations are already linearized since the Weyl scalars
$\Psi_{0}$ and $\Psi_{4}$ vanish in the equilibrium state, and then
they are to be considered first order perturbations. The other
quantities take their unperturbed values given in
Eqs. \eqref{vec-l}-\eqref{curvs}.

The background symmetries, together with the perturbation equations
\eqref{fund1} and \eqref{fund2}, indicate that each one of the
perturbation functions $\Psi_{0}$ and $\Psi_{4}$ can be written as a
product of four different functions depending upon only one of the
coordinates $t$, $r$, $\varphi$, and $z$. Moreover, due to the
wavelike character of the equations, the dependence of these functions
on the coordinates $t$, $\varphi$ and $z$ is of the form
$\mbox{exp}[i(\omega t+k_{\varphi} \varphi+k_{z}z)]$, where $\omega$
is the frequency, and $k_{\varphi}$ and $k_{z}$ are wave numbers which
may assume discrete or continuous values, depending on whether
$\varphi$ and $z$ are compact or noncompact coordinates,
respectively. The radial functions, in turn, are conveniently written as
\begin{equation*}
\Psi_{0}(r)=r^{-1}f^{-2}Y_{+2}(r),
\qquad\quad\Psi_{4}(r)=r^{-1}Y_{-2}(r),
\end{equation*}
where the subscripts $\pm 2$ are related to the
spin weight and to the conformal weight of the
Weyl scalars $\Psi_{0}$ and $\Psi_{4}$.

The resulting equations for $Y_{\pm 2}$ take the following standard form: 
\begin{equation}
\Lambda^{2}Y_{\pm 2}+P\Lambda_{\mp}Y_{\pm 2}-QY_{\pm 2}=0,
\label{eq-desejada1}
\end{equation}
where we introduced the differential operators
\begin{equation*}
\Lambda_{\pm}=\frac{d}{dr_{\ast}}\pm i\omega ,
\qquad\quad\Lambda^{2}=\frac{d^{\,2}}{dr_{\ast}^{2}}+\omega^{2}.
\end{equation*}
Here $r_{\ast}$ is the tortoise coordinate, defined in 
such a way that $dr/dr_{\ast}=f(r)$, and
\begin{equation*}
P=\frac{d}{dr_{\ast}}\ln\left(\frac{r^{4}}{f^{2}}\right),
\qquad\quad Q=\frac{f}{r^{3}}(k^{2}r+12M),
\end{equation*}
with $k^2=k_{\varphi}^2+k_{z}^2$.

The Chandrasekhar transformation theory \cite{chan1} can now be used
to transform Eqs. \eqref{eq-desejada1} into one-dimensional
Schr\"odingerlike equations of the form
\begin{equation}
\Lambda^{2}Z^{ (\pm)}=V^{ (\pm)}
Z^{ (\pm)}.
\label{wave}
\end{equation}
This is done by means of the substitutions
\begin{gather*}
Y_{+2}=V^{(\pm)}Z^{ (\pm)}
+(W^{ (\pm)}+
2i\omega)\Lambda_{+}Z^{ (\pm)},\\
Y_{-2}=V^{ (\pm)}Z^{ (\pm)}
+(W^{ (\pm)}-
2i\omega)\Lambda_{-}Z^{ (\pm)},
\end{gather*}
whose inverse transformation gets the form
\begin{gather*}
K^{ (\pm)}Z^{ (\pm)}=
\frac{r^{4}}{f^{2}}QY_{+2}-\frac{r^{4}}{f^{2}}(W^{ 
(\pm)}+2i\omega)\Lambda_{-}Y_{+2},\\
K^{ (\mp)}Z^{ (\pm)}=
\frac{r^{4}}{f^{2}}QY_{-2}-\frac{r^{4}}{f^{2}}(W^{ 
(\pm)}-2i\omega)\Lambda_{+}Y_{-2},
\end{gather*}
where 
\begin{equation}
K^{ (\pm)}=k^4\pm 24i\omega M.
\label{kappapm}
\end{equation}
In the foregoing equations, the potential functions
$V^{ (\pm)}$ are given by
\begin{equation}
V^{ (+)}=\frac{f}{r^{3}}
\left[\frac{576M^{3}+12k^{4}Mr^{2}+k^{6}r^{3}
+144M^{2}r(k^{2}+2r^{2})}{(k^{2}r+12M)^{2}}\right],
\label{pot-polar}
\end{equation}
\begin{equation}
V^{ (-)}=\frac{f}{r^{3}}(k^{2}r-12M),
\label{pot-axial}
\end{equation}
while the auxiliary functions $W^{ (\pm)}$
are, respectively,
\begin{equation*}
W^{ (+)}=-\frac{12M(2r^{3}+k^{2}r+4M)}{r^{2}(k^{2}r+12M)},
\qquad\quad W^{ (-)}=-\frac{12M}{r^{2}}.
\end{equation*}

The meaning of the signs $(\pm)$ comes from the metric approach to the
perturbation problem, as it was performed in Ref. \cite{car7}. They
are related to the parity property of the perturbed metric functions
under the exchange $\varphi\rightarrow-\varphi$ (or
$z\rightarrow-z$). The variable $Z^{ (+)}$
represents the polar (even) perturbations, while $Z^{ (-)}$
represents the axial (odd) ones.

In a certain sense, the wave equations \eqref{wave} can be thought of as
a generalization to those found by Cardoso and Lemos \cite{car7}. Here
the perturbations depend on both coordinates $\varphi$ and $z$, while
in the aforementioned work the analysis was restricted to
$z$-independent metric variations. This is equivalent to the vanishing
of the wave number $k_{z}$, resulting in $k^2=k_{\varphi}^2$. However,
the plane symmetry of the background spacetime renders possible to
find new coordinates ($\varphi '$, $z'$) in such a way that one of the
wave numbers, $k_\varphi'$ or $k_z'$, is zero. That is to say, it is
always possible to find a coordinate system in which the perturbations
depend only on two space coordinates ($r$ and $\varphi$, or $r$ and
$z$).

Another result derived from the above analysis is related to the
Starobinsky constant \cite{staro},
$|\mathscr{C}|^2=K^{ (+)}K^{
(-)}$, where $K^{ (\pm)}$ are given by
Eq. \eqref{kappapm}. The solutions of $|\mathscr{C}|^2=0$ correspond
to algebraically special frequencies in the sense of Chandrasekhar
\cite{chan2}. As we shall see in Sec. \ref{resultados}, one of such
solutions, given by
\begin{equation}
\omega_{a}=i\frac{k^4}{24M}=i\frac{k^4}{6r_{+}^{3}},
\label{algebraicfreq}
\end{equation}
appears in the numerical results in connection to the purely
damped QNMs. The properties of the axial perturbations
related to these algebraically special modes are analyzed in
some detail in Sec. \ref{algebrico}.

\section{QNM analytical properties}
\label{qnm}

To solve the wave equations \eqref{wave} we must impose boundary
conditions on the possible solutions. First, however, consider the
case of the asymptotically flat Schwarzschild spacetime.
Gravitational perturbations lead to Schr\"odingerlike
equations whose potentials vanish at infinity and at the event
horizon. Hence, close to these extremities, the solutions are plane
waves, $Z\rightarrow e^{\pm i\omega r_{\ast}}$, where the $r_{\ast}$
coordinate ranges from $-\infty$ to $+\infty$. QNMs are then defined
as solutions which are purely ingoing waves at the horizon, $Z\rightarrow
{\rm e}^{+i\omega r_{\ast}}$, and purely outgoing waves at infinity,
$Z\rightarrow {\rm e}^{-i\omega r_{\ast}}$. These requirements are
satisfied only for a discrete set of complex $\omega$ called the
quasinormal frequencies.

In asymptotically AdS spacetimes, we keep the requirement of purely
ingoing waves at the horizon. In the meantime, since now the tortoise
coordinate $r_{\ast}$ takes a finite value at spatial infinity, the
boundary condition at infinity must be changed. This can be Dirichlet,
Neumann, or Robin boundary condition, depending on whether the field,
its derivative, or a combination of both vanishes, respectively. Here,
as in most of the literature, we adopt Dirichlet boundary
conditions, since these lead to energy conservation in the field
theory on AdS space \cite{avis} (for alternative boundary conditions
see Ref. \cite{moss}).

Having defined the QNMs, we show below that the imaginary part of the
QN frequencies is positive for non-negative potentials.  The
demonstration follows the one carried out in the Schwarzschild-AdS case
by Cardoso and Lemos \cite{car1}. Initially, we exchange the variable
$Z$ by $\phi=e^{-i\omega r_{\ast}}Z$, so that Eq. \eqref{wave} becomes
\begin{equation}
f\frac{d^{\,2}\phi}{dr^{2}}+\left(\frac{df}{dr}+
2i\omega\right)\frac{d\phi}{dr}-\frac{V}{f}\phi=0,
\label{fund-r}
\end{equation}
where $Z$, $\phi$ and $V$ stand for both perturbation types,
the polar $(Z^{ (+)},\;\phi^{ (+)},\;
V^{ (+)})$ and the axial ones 
$(Z^{ (-)},\;
\phi^{ (-)},\;V^{ (-)})$.

Near the horizon and at infinity, the solutions $\phi$
satisfying the QNM boundary conditions are given, respectively, by
\begin{equation}
\phi_{r_{+}}=A\frac{(C_{1}/3r_{+})^{i\omega/3r_{+}}}{\Gamma[1+
(2i\omega/3r_{+})]},
\label{assint1}
\end{equation}
\begin{equation}
\phi_{\infty}=B\,{\rm e}^{i\omega/r}\sin
\left(\sqrt{\omega^{2}-C_{2}}/r\right),
\label{assint2}
\end{equation}
where $A$ and $B$ are arbitrary constants, and 
$C_{1}=V(r_{+})/f(r_{+})$ and $C_{2}=V(\infty)$ are
definite constants whose values depend on the
perturbation type.

Now we multiply Eq. \eqref{fund-r} by $\phi^{\ast}$ and integrate by
parts using the asymptotic solutions \eqref{assint1} and
\eqref{assint2}. After some algebra we obtain
\begin{equation}
\int^{\infty}_{r_{+}}\bigg[f\left|\frac{d\phi}{dr}\right|^{2}+
\frac{V}{f}|\phi|^{2}\bigg]dr=
\frac{|\omega|^{2}|\phi(r_{+})|^{2}}{\mbox{Im}(\omega)}.
\label{int-final}
\end{equation}
Since $f(r)$ assumes only positive values outside of the black hole,
the sign of the quantity on the left hand side of Eq.
\eqref{int-final} depends crucially on the sign of the potential
$V(r)$. For non-negative $V$, it follows that the imaginary part of
the QN frequencies is positive, $\Im(\omega)> 0$. Moreover, it is seen
from Eq. \eqref{pot-polar} that $V^{ (+)}$ is
positive in the region $r_{+}<r<\infty$ for any $k$ and $r_{+}$, and
then we conclude that the plane-symmetric AdS black holes are stable
against polar perturbations. The same is true for axial perturbations
as far as the wave numbers are restricted by $k\geq\sqrt{3}r_{+}$. On
the other hand, for $k<\sqrt{3}r_{+}$, the potential
$V^{ (-)}$ takes negative values and the theorem
cannot be applied.

Another interesting property of frequencies associated to
perturbations of a plane-symmetric black hole comes from the
particular symmetry of such a background. This symmetry allows us to
rescale $\omega$, $k$ and $r_{+}$ by a pure scale transformation of
coordinates $t=a{t'}$, $\varphi=a{\varphi'}$, $z=a {z'}$, and
$r={r'}/a$, where $a$ is a constant \cite{horo}. It implies that the
frequency is a first degree homogeneous function of its parameters:
$\omega(ar_{+},ak)= a\omega(r_{+},k)$. In particular, we can take
$a=k_{0}/k$ so that
\begin{equation}
\omega(r_{+},k)=\frac{k}{k_{0}}\,
\omega\left(\frac{k_{0}}{k}\,r_{+},k_{0}\right).
\label{relacao-k}
\end{equation}
Then, after computing the QN frequencies for a given $k$, e. g.,
for $k=k_{0}$, and for different values of $r_{+}$, we can obtain
$\omega(r_{+},k)$ by using the previous equation, and the values of
$\omega$ for a fixed wave number
$k$ are naturally divided into three different regimes: large
($r_{+}\gg k$), intermediate ($r_{+}\sim k$) and small ($r_{+}\ll
k$) black holes. In the large (small) black hole limit, $k$ ($r_{+}$)
is negligibly small and it is expected that $\omega\propto r_{+}$
($\omega\propto k$).

We are now in a position to solve the perturbation equations
\eqref{wave} and to find the values of the QN frequencies. This task,
however, is more easily done through numerical procedures. Therefore,
it is appropriate summarizing here the method used for such a purpose.

\section{The Horowitz-Hubeny Method}
\label{metodo}

In the last decades a series of methods has been developed to
numerically calculate the QNMs of black holes. Here we briefly
describe one of these methods, used in Ref. \cite{horo}, which is of
particular interest for the present work since it is suited for
asymptotically AdS spacetimes. In order to apply such a method we
rewrite here the perturbation equations in
the appropriate form.

By introducing the independent variable $x=1/r$,
Eq. \eqref{fund-r} becomes
\begin{equation}
(x^{3}-x_{+}^{3})\frac{d^{\,2}\phi}{dx^{2}}+(3x^{2}+2i\omega x_{+}^{3})
\frac{d\phi}{dx}+\widetilde{V}\phi=0,
\label{fund-x}
\end{equation}
where $x_{+}=1/r_{+}$ and $\widetilde{V}(x)= x_{+}^{3}V/f x^{2}$, with
$V$ and $f$ now being considered as functions of $x$.  The foregoing
equation has a regular singular point ($x=x_{+}$) in the region of
interest, $0\leq x\leq x_{+}$. Then, by Fuchs theorem, any solution of
\eqref{fund-x} can be written as a Fr\"obenius series
\cite{butkov}. The behavior of the solutions near the horizon may be
explored by writing $\phi(x)=(x-x_{+})^{\alpha}$, where $\alpha$ is a
constant to be determined. By substituting this expression into
Eq. \eqref{fund-x} and keeping terms only to the leading order, it
follows that $x_{+}^{2}\alpha(3\alpha+2i\omega x_{+})=0$, which has the
roots $\alpha=0$ and $\alpha=-2i\omega/3r_{+}$. These roots correspond
exactly to the ingoing and outgoing waves at the horizon, respectively
\cite{horo}. Since we want only ingoing modes, we take $\alpha=0$ and
look for solutions of the form
\begin{equation}
\phi(x)=\sum_{m=0}^{\infty}a_{m}(x-x_{+})^{m}.
\label{serie}
\end{equation}
Substituting Eq. \eqref{serie} into \eqref{fund-x}, we find
expressions for the $a_{m}$'s in terms of the frequency $\omega$ (and
of the parameters $k$ and $x_{+}$). Then, imposing the boundary
condition at infinity, $\phi(0)=0$, we obtain
\begin{equation}
\sum_{m=0}^{\infty}a_{m}(\omega)\,\left(-x_{+}\right)^{m}=0.
\label{eq-serie}
\end{equation}
Thus, the problem of finding QN frequencies has been reduced to that
of obtaining the roots of the infinite-order polynomial equation
\eqref{eq-serie}. Of course, we cannot carry out numerically the full
sum in expression \eqref{eq-serie}. In effect we truncate the series
after a large number of terms and look for the zeros of this partial
sum. The precision of the results is then verified by computing the
relative variation between the zeros as we perform higher partial
sums. We stop our search once we have a $6$ decimal digit
precision. For each pair of $k$ and $r_{+}$, we find a set of QN
frequencies (out of the infinity of possible frequencies). These
frequencies are labeled as usual with the principal quantum number $n$
and ordered in a set beginning with the roots with the lowest imaginary
parts.

\section{Numerical results and analysis}
\label{resultados}

We use the Horowitz-Hubeny procedure described above to
compute the gravitational QNMs of plane-symmetric AdS black
holes. As it happens to other AdS black holes, there are
{\it{purely damped modes}}, whose frequencies are pure imaginary numbers, 
in addition to the {\it{ordinary modes}}, the frequencies of which also have
nonzero real parts which are responsible for the oscillatory behavior of the
perturbations.

As mentioned earlier, the numerical analysis is divided
into three different regimes according to the ratio
$r_{+}/k$ values. That is to say, we have (i) large black
holes when the ratio $r_{+}/k$ is large compared to unity;
(ii) intermediate black holes if $r_{+}/k$ is of the
order of unity; and (iii) small black holes when $r_{+}/k$
is small compared to unity.
 
It is also worth mentioning here that some numerical analysis
of the gravitational QNMs was done in Ref. \cite{car7}. However,
such a study was restricted mainly to the large black hole
regime for a specific value of $k$, without generalizations for a
generic wave number. In what follows, we perform a more detailed analysis,
including small and intermediate black holes, and extend implicitly
the numerical results to all values of $k$.
We also search the QN frequency values in the limit
of highly damped overtones.

\subsection{Ordinary quasinormal modes}
\label{ordinario}

To begin with, we discuss the case of regular (ordinary) QNMs for
which the frequencies have nonvanishing real and imaginary parts,
$\omega=\omega_{r}+i\omega_{i}$. We list some numerical results of the
lowest quasinormal frequencies for $k=2$ and some selected values of
$r_{+}$. The computed quantities include both the axial and polar
perturbations, and the three regimes defined above. The data for
large, intermediate, and small black holes are shown, respectively,
in Tables \ref{tlarge}, \ref{tmedium}, and \ref{tsmall}.

\begin{table}
\begin{tabular}{|c|c|c|c|c|c|}
\hline\hline
& \multicolumn{2}{|c|}{Axial} & \multicolumn{2}{|c|}
{Polar}\\
\cline{2-5}
$\quad r_{+}\quad $ & $\qquad\omega_{r}\qquad $ &
$\qquad\omega_{i}\qquad $ & $\qquad\omega_{r}\qquad $ &
$\qquad\omega_{i}\qquad $\\
\hline
1000 & 1849.42 & 2663.84 & 1849.45 & 2663.85\\
500 & 924.710 & 1331.92 & 924.727 & 1331.92\\
100 & 184.948 & 266.384 & 184.963 & 266.351\\
50 & 92.4820 & 133.189 & 92.5134 & 133.124\\
10 & 18.5492 & 26.6228 & 18.6960 & 26.2961\\
5 & 9.35491 & 13.2880 & 9.62439 & 12.6329\\
\hline\hline
\end{tabular}
\centering
\caption{The lowest ordinary QN frequencies for $k=2$ and several
values of $r_{+}$ in the large black hole regime.}
\label{tlarge}
\end{table}

\begin{table}
\begin{tabular}{|c|c|c|c|c|c|}
\hline\hline
& \multicolumn{2}{|c|}{Axial} & \multicolumn{2}{|c|}
{Polar}\\
\cline{2-5}
$\quad r_{+}\quad $ & $\qquad\omega_{r}\qquad $ &
$\qquad\omega_{i}\qquad $ & $\qquad\omega_{r}\qquad $ &
$\qquad\omega_{i}\qquad $\\
\hline
2 & 3.92888 & 5.24698 & 4.06031 & 3.84425\\
1.6 & 3.21225 & 4.15458 & 3.21731 & 2.78751\\
1.2 & 2.44666 & 2.99891 & 2.52608 & 1.92268\\
1 & 2.04735 & 2.21550 & 2.30526 & 1.55218\\
0.8 & 2.05189 & 1.41550 & 2.18361 & 1.19971\\
0.4 & 2.09860 & 0.520025 & 2.10247 & 0.511896\\
\hline\hline
\end{tabular}
\centering
\caption{The lowest ordinary QN frequencies for $k=2$ and several
values of $r_{+}$ in the intermediate size black hole regime.}
\label{tmedium}
\end{table}

\begin{table}
\begin{tabular}{|c|c|c|c|c|c|}
\hline\hline
& \multicolumn{2}{|c|}{Axial} & \multicolumn{2}{|c|}
{Polar}\\
\cline{2-5}
$\quad r_{+}\quad $ & $\qquad\omega_{r}\qquad $ &
$\qquad\omega_{i}\qquad $ & $\qquad\omega_{r}\qquad $ &
$\qquad\omega_{i}\qquad $\\
\hline
1/10 & 2.02727 & 0.0916071 & 2.02727 & 0.0915965\\
1/12 & 2.02223 & 0.0733694 & 2.02223 & 0.0733649\\
1/14 & 2.01866 & 0.0608423 & 2.01866 & 0.0608402\\
1/16 & 2.01601 & 0.0517477 & 2.01601 & 0.0517466\\
1/18 & 2.01397 & 0.0448789 & 2.01397 & 0.0448783\\
1/20 & 2.01236 & 0.0395123 & 2.01236 & 0.0395118\\
\hline\hline
\end{tabular}
\centering
\caption{The lowest ordinary QN frequencies for $k=2$ and several
values of $r_{+}$ in the small black hole regime.}
\label{tsmall}
\end{table}

The first feature to be noticed is that, contrary to the
asymptotically flat space case, the axial and polar perturbations no
longer have the same spectra. In fact, the isospectrality is only
restored in the large and small black hole limits (see also
Refs. \cite{car1,bert1}).

The displayed numerical results also allow us to obtain
the $k$ dependence of the modes on basis of the homogeneity
property of the function $\omega(r_{+},k)$, as represented by
Eq. \eqref{relacao-k}. More specifically, the $k\neq 0$
associated frequency can be found by means of the relation
\begin{equation}
\omega(r_{+},k)= \frac{k}{2}\;\omega\left(2\,\frac{r_{+}}{k},2\right).
\label{relacao-2}
\end{equation}

For large black holes, our results are in perfect agreement with those
found in Ref. \cite{car7}. Axial and polar perturbations have real and
imaginary parts of the fundamental frequencies given approximately by
$\omega_{r}=1.85 r_{+}$ and $\omega_{i}=2.66 r_{+}$. According to
Eq. \eqref{relacao-2}, these linear fits are wave-number
independent. They are also equal to those from the large
Schwarzschild-AdS black hole \cite{car1}, as should be expected,
since a spherical system approaches a plane one as the radius goes to
infinity.

\begin{figure}[ht]
\centering\epsfig{file=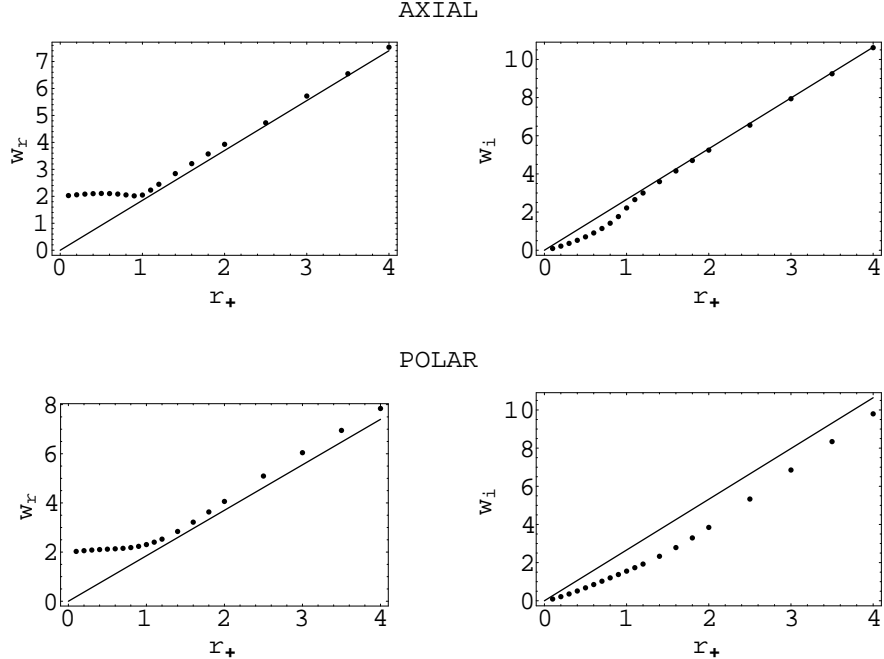}
\caption{The frequency of ordinary QNMs for axial and polar
perturbations, $\omega=\omega_{r}+i\omega_{i}$, as a function of the
horizon radius $r_{+}$ for small and intermediate black
holes. The dots represent numerical results, and the solid lines are
linear fits. On the left are the real parts ($\omega_r$) with the linear fit
$1.85 r_{+}$, on the right, the imaginary parts ($\omega_{i}$), with the
fit line $2.66r_+$.}
\label{fordinario}
\end{figure}

For small and intermediate black holes, the quasinormal
frequencies do not scale with the horizon radius. This is clearly
shown in Fig. \ref{fordinario}, where we plot the lowest axial and
polar frequencies as functions of $r_{+}$, for $k=2$ and $0.1\leq
r_{+}\leq 4$ . The exact way the curve $\omega(r_{+})$ deviates from
the linear behavior depends on the perturbation parity. Within a
certain accuracy, the fundamental frequency starts deviating from the
line $\omega=(1.85+i2.66)r_{+}$ around $r_{+}=k/2$ for axial modes,
and around $r_{+}=5k/2$ for polar modes. As the ratio $r_{+}/k$ goes
to zero, we have $\omega_{r}\rightarrow k$ and $\omega_{i} \rightarrow
0$, independently of the perturbation type.
In fact, the dependence of $\omega$ upon $k$ in the limit of small black holes
may be obtained using Eq. \eqref{relacao-k} which gives $\omega(r_{+},k)
\approx\omega\left(0,k_{0}\right){k}/{k_{0}}\, = const.\times
k\, .$ The numerical analysis shows that for the fundamental mode
the proportionality constant is equal to unity.
The relation of this result to the pure AdS modes is investigated
in Appendix A.

\subsection{Purely damped modes}
\label{amortecido}

Axial perturbations of AdS black holes are known to have special
quasinormal modes which are purely damped. In the present case, the
values of the associated frequencies, $\omega=i\omega_{s}$, are listed
in Table \ref{tpuro} for $k=2$ and some selected values of
$r_{+}$. Once again, the homogeneity property of $\omega(r_{+},k)$
allows us to obtain the $k$ dependence of the frequency on the basis of
relation \eqref{relacao-2}.

\begin{table}
\begin{tabular}{|c|c|c||c|c||c|c|c|}
\hline\hline
$r_{+}\;$ & $\quad\omega_{s}\quad$ &
$\quad4/3r_{+}\quad$ & $r_{+}\;$ &
$\quad\omega_{s}\quad$ & $r_{+} \;$ &
$\quad\omega_{s}\quad$ & $ 8/3r_{+}^{3}$\\
\hline
100 & 0.0133342 & 0.0133333 & 2.0 & 0.733037 & 0.9 & 3.66330 & 3.65798\\
75 & 0.0177794 & 0.0177778 & 1.8 & 0.835965 & 0.7 & 7.77447 & 7.77454\\
50 & 0.0266704 & 0.0266667 & 1.6 & 0.978431 & 0.5 & 21.3333 & 21.3333\\
25 & 0.0533639 & 0.0533333 & 1.4 & 1.19459 & 0.3 & 98.7654 & 98.7654\\
10 & 0.133786 & 0.133333 & 1.2 & 1.58723 & 0.2 & 333.333 & 333.333\\
5 & 0.270361 & 0.266667 & 1.0 & 2.64636 & 0.1 & 2666.67 & 2666.67\\
\hline\hline
\end{tabular}
\centering
\caption{The pure imaginary QN frequency for $k=2$ and some selected
values of $r_{+}$. The formula $4/3r_{+}$ is a fit valid for large
black holes, while $8/3r_{+}^{3}$ is the magnitude of the
algebraically special frequency $\omega_{a}$.}
\label{tpuro}
\end{table}

For large black holes, our numerical results agree with those
presented in Ref. \cite{car7}. To a remarkable accuracy, the $k=2$
pure imaginary mode is well described by the formula
$\omega_{s}=4/3r_{+}$. The extension of this fit to any $k$ leads to
$\omega_{s}=k^2/3r_{+}$. A similar result was already obtained for the
Schwarzschild-AdS spacetime by Cardoso {\it{et al}} \cite{car5}. They
found $\omega_{s}=(l-1)(l+2)/3r_{+}$, where $l$ is the angular
momentum of the perturbation. The above two formulas for $\omega_s$
are just fits to the numerical data, but we were able to find an exact
analytical proof for both of them by expanding $\omega_{s}$ in a power
series of $1/r_{+}$. This study is the subject of the next section.

\begin{figure}
\centering\epsfig{file=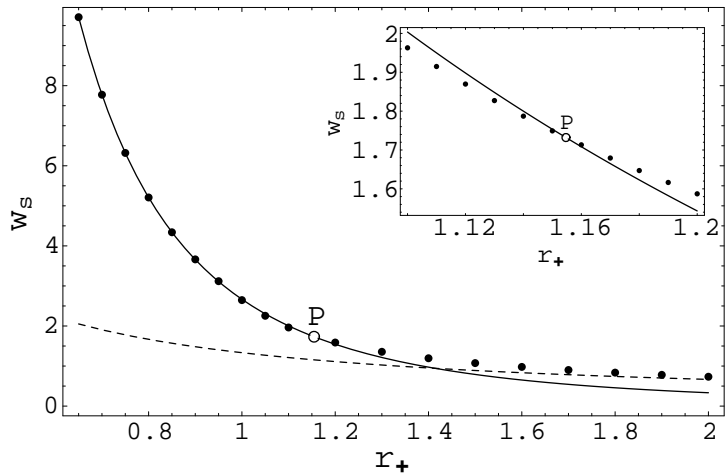}
\caption{The frequency $\omega_{s}$ of the $k=2$ purely damped mode
for intermediate and small black holes as a function of the
horizon radius (the dots represent numerical results). The dashed
line is $\omega_{s}=4/3r_{+}$, and the solid line is the algebraically
special frequency $|\omega_{a}|=8/3r_{+}^{3}$. At point
$\mbox{P}=(2/\sqrt{3},\sqrt{3})$, the algebraically special frequency
line crosses the purely damped QN frequency.}
\label{fpuro}
\end{figure}

For intermediate size black holes, the frequency $\omega_{s}$
increases faster than $k^2/3r_{+}$ as the horizon radius
decreases (see Fig. \ref{fpuro}).
Another special feature of this regime is related to the
behavior of the axial potential $V^{ (-)}$ as a
function of the coordinate $r$. As we go from larger to smaller
horizon radii in the plane $r_{+}\times k$, the straight line
$k=\sqrt{3}r_{+}$ is the boundary on which the axial potential
\eqref{pot-axial} becomes a non-negative function in the region
$r_{+}\leq r<\infty$. In other words, $V^{
(-)}(r)\geq 0$ at all points outside of a black hole perturbed by an
axial mode with $k\geq \sqrt{3}r_{+}$. A consequence of this for the
numerical calculations is that, at $r_{+}=k/\sqrt{3}$, our
root-searching program based on the Horowitz-Hubeny method fails to
find a pure imaginary QN frequency. The problem occurs when $r_+$
approaches the point where the function $\omega=i\omega_{s}(r_+)$
would intercept the algebraically special frequency
$\omega_{a}=ik^{4}/6r_{+}^{3}$. This is indicated by the point $\mbox{P}$
in Fig. \ref{fpuro}, where we plot $\omega_{s}$ as a function of $r_{+}$
for $k=2$. The numerical data show that $\omega_s >|\omega_a|$ for all
$r_+> k/\sqrt{3\,}$, and $\omega_s <|\omega_a|$ for all
$r_+< k/\sqrt{3\,}$.

For small black holes, the purely damped frequency $i\omega_s$ approaches
the algebraically special frequency $\omega_{a}$ from below.
In fact, the difference between these frequencies is smaller
than $10^{-6}$ for $r_{+}/k=0.3$ and decreases as $r_{+}/k\rightarrow 0$.
This conclusion has again its
spherical counterpart, as it can be seen from the numerical results
found in Ref. \cite{car5}. More details on the algebraically special
frequency and on its non-QN character are given in
Sec. \ref{algebrico}.

\subsection{Highly damped modes}
\label{higher}

For the moment, we present only the computed frequency
values of the lowest QNMs. They are the most important
modes for the AdS/CFT correspondence, since these give
the thermalization time scale on the boundary
field theory. However, as soon as one is interested
in Hod's conjecture \cite{hod}, which relates the
classical QN frequencies with the quantum
spectrum of black holes, the most relevant modes
are the highly damped QNMs. Indeed, according to
such a conjecture, the real part of the QN frequencies,
in the limit of high overtones, gives information about
the quantum of the black hole area. In the following,
we present the QN frequency values for modes in the
regime of high principal quantum numbers $n$.
Since we can distinguish the purely damped modes
as belonging to a special family, we simply label
them with $n=0$, and begin to label the fundamental
ordinary modes with $n=1$.

\begin{table}
\begin{tabular}{|c|c|c|c|c|}
\hline\hline
& \multicolumn{2}{|c|}{Axial} & \multicolumn{2}{|c|}
{Polar}  \\
\cline{2-5}
$\quad n \quad $ & $\qquad\;\;\omega_{r}\qquad $ &
$\qquad\omega_{i}\qquad $ & $\qquad\;\;\omega_{r}\qquad $ &
$\qquad\omega_{i}\qquad $\\
\hline
1 & 184.948 & 266.384 & 184.963 & 266.351 \\
2 & 316.130 & 491.641 & 316.159 & 491.583\\
3 & 446.438 & 716.753 &  446.482 & 716.671\\
4 & 576.528 & 941.807 & 576.586 & 941.700\\
5 & 706.536 & 1166.84 & 706.608 & 1166.71\\
20 & 2655.34 & 4541.91 & 2655.62 & 4541.40\\
40 & 5253.44 & 9041.92 & 5254.09 & 9041.02\\
\hline\hline
\end{tabular}
\centering
\caption{QNMs corresponding to $k=2$ axial and polar perturbations
of a large plane-symmetric AdS black hole, with $r_{+}=100$.}
\label{highlydampedlarge}
\end{table}

To study the QNMs asymptotic behavior, we
computed the first 40 ordinary QN frequencies
associated to $k=2$ axial and polar perturbations
of plane-symmetric AdS black holes. A few representative results for
some modes of large, intermediate, and small event horizons
are displayed in Tables \ref{highlydampedlarge},
\ref{highlydampedmedium}, and \ref{highlydampedsmall}, respectively.

\begin{table}
\begin{tabular}{|c|c|c|c|c|}
\hline\hline
& \multicolumn{2}{|c|}{Axial} & \multicolumn{2}{|c|}
{Polar}  \\
\cline{2-5}
$\quad n \quad $ & $\qquad\;\;\omega_{r}\qquad $ &
$\qquad\omega_{i}\qquad $ & $\qquad\;\;\omega_{r}\qquad $ &
$\qquad\omega_{i}\qquad $\\
\hline
1 & 2.04735 & 2.21550 & 2.30526 & 1.55218\\
2 & 3.34678 & 4.74580 & 3.26137 & 3.72163\\
3 & 4.65907 & 7.05251 & 4.39306 & 5.96477\\
4 & 5.95653 & 9.32552 & 5.59513 & 8.21840\\
5 & 7.24999 & 11.5884 & 6.83055 & 10.4733\\
20 & 26.6726 & 45.3810 & 26.0793 & 44.2560\\
40 & 52.6238 & 90.3921 & 44.2193 & 75.7648\\
\hline\hline
\end{tabular}
\centering
\caption{QNMs corresponding to $k=2$ axial and polar perturbations
of an intermediate size plane-symmetric AdS black hole, with $r_{+}=1$.}
\label{highlydampedmedium}
\end{table}

\begin{table}
\begin{tabular}{|c|c|c|c|c|}
\hline\hline
& \multicolumn{2}{|c|}{Axial} & \multicolumn{2}{|c|}
{Polar}  \\
\cline{2-5}
$\quad n \quad $ & $\qquad\;\;\omega_{r}\qquad $ &
$\qquad\omega_{i}\qquad $ & $\qquad\;\;\omega_{r}\qquad $ &
$\qquad\omega_{i}\qquad $\\
\hline
1 & 2.05667 & 0.214983 & 2.05678 & 0.214687\\
2 & 2.17823 & 0.576255 & 2.17845 & 0.575676 \\
3 & 2.32373 & 0.974834 & 2.32406 & 0.974002\\
4 & 2.48740 & 1.39433  & 2.48783 & 1.39325\\
5 & 2.66559 & 1.82658  & 2.66613 & 1.82528\\
20 & 6.04295 & 8.60462 & 6.04536 & 8.60001\\
40 & 11.0316 & 17.6657 & 11.0366 & 17.6567\\
\hline\hline
\end{tabular}
\centering
\caption{QNMs corresponding to $k=2$ axial and polar perturbations
of a small plane-symmetric AdS black hole, with $r_{+}=0.2$.}
\label{highlydampedsmall}
\end{table}

On the basis of this numerical data, we obtained linear fits by taking
overtones from $n=21$ to $n=40$ and employing the
discrete least squares approximation method.
Together with the coefficients of the linear approximating functions,
we calculated the error involved in each fit process and established
$1\%$ as the upper bound error. This means that, in the regime of high
overtones ($n\rightarrow\infty$), the
QN frequencies are evenly spaced with real and imaginary parts given
approximately by linear functions as described below.

The linear fits depend upon the size of the horizon radius, and 
are given by functions of the form 
\begin{equation}
\omega_n = \alpha  +\beta\, n\,  ,
\end{equation}
where the coefficients $\alpha$ and $\beta$ refer to both
the real and the imaginary parts of the frequency, and are given 
in table \ref{highovertonesfits}. Their dependence upon the black hole
parameter $r_+$ and on the wave number $k$ is governed by
Eq. \eqref{relacao-k}.

\begin{table}
\begin{tabular}{|c|c|c|c|c|}
\hline\hline
& \multicolumn{2}{|c|}{Axial} & \multicolumn{2}{|c|}
{Polar}  \\
\cline{2-5}
$\quad r_+ \quad $ & $\qquad\alpha\qquad $ &
$\qquad\beta\qquad $ & $\qquad\alpha\qquad $ &
$\qquad\beta\qquad $\\
\hline
100 &  $57.2447+41.9027i$ &   $129.905 +225.000i$  &  $57.0927 + 41.7360i$
& $129.925 + 224.982i$ \\
1  & $0.717273 +0.371668i $ & $1.29763 + 2.25053i $ & $-7.58881-14.2636i $
& $ 1.29504+ 2.25074i $ \\
0.2 & $1.02468-0.451644i $ & $0.249900+0.452989i $ & $ 1.02451-0.451813i $
&$ 0.250029+0.452767i  $ \\
\hline\hline
\end{tabular}
\centering
\caption{Linear fits for the frequencies of high overtones of
axial and polar perturbations for black holes
with $r_+=100$, $r_+=1$, and $r_{+}=0.2$.}
\label{highovertonesfits}
\end{table}

As should be expected, the results found for
large black holes reproduce those obtained in Ref.
\cite{car5} for Schwarzschild-AdS black holes with a
large event horizon. A special feature of the regime is that
axial and polar perturbations have essentially the same QNM
spectra, which appear in the first line of Table \ref{highovertonesfits}.
In the other regimes, however,
the linear functions are very different from those
of the spherical case.

The angular coefficients of the linear functions seem to present a kind of
universal behavior. In fact, as seen from the $\beta$ coefficients in
Table \ref{highovertonesfits}, the results for both of the perturbation
types, and for each horizon radius, are very close to each other.
The spacings between the frequencies per unit horizon
radius of consecutive modes are essentially independent of the
black hole size. This is seen more clearly in Table
\ref{highovertonesspacing} where the values of the ratio $(\omega_{n+1}
-\omega_{n})/r_+$, calculated for each regime, are shown.
The data confirm that such a ratio, for the calculated wave number
$k=2$, is approximately constant.
Additionally, using this result together with Eq. \eqref{relacao-2} it may
be shown that, in the high overtones regime, the frequency per unit horizon
radius spacing is not only independent of the black hole size and of the
perturbation parity, but it is also independent of the wave-number value.
The real parts for small black holes are about $4\%$ lower than for
intermediate and large black holes. We notice that this difference
decreases as we search for overtones with principal quantum numbers much
larger than $n=40$, but we have not investigated this more precisely
because the numerical calculations become very slow for overtones with
$n$ significantly larger than $40$.

\begin{table}
\begin{tabular}{|c|c|c|}
\hline\hline
& \multicolumn{2}{|c|}
{$\displaystyle {(\omega_{n+1}-\omega_{n})/ {r_{+}}}$}\\
\hline
$\quad r_+\quad$& Axial & Polar \\
\hline
100 & $\quad 1.29905+2.25000i\quad $& $\quad1.29925+2.24982i\quad$ \\
  1  & $\quad 1.29763+2.25053i\quad$        &$\quad 1.29504+2.25074i\quad$\\
0.2 &$ \quad 1.24950+2.26495i\quad $  & $\quad 1.25015+2.26384i \quad $ \\
 \hline\hline
\end{tabular}
\centering
\caption{The fractional spacing of frequencies among consecutive
high overtones of axial and polar perturbations
of black holes with $r_+=100$, $r_+=1$, and $r_{+}=0.2$.}
\label{highovertonesspacing}
\end{table}

\section{The large black hole limit of
the purely damped QNM}
\label{limite}

Although numerical procedures need to be used to compute the QN
frequencies, it is possible to study analytically the special
case of axial perturbations of very large black holes.
Here we show in some detail the process of finding an asymptotic
expression for the frequency of the purely damped modes.
The analysis is performed in such a way that it holds for both
the plane-symmetric and the Schwarzschild-AdS spacetimes.

Our starting point is the Horowitz-Hubeny approach described in
Sec. \ref{metodo}. There, the QN frequencies of axial perturbations
are associated to solutions of the equation
\begin{equation}
\sum_{m=0}^{\infty}a_{m}(\omega)\,\left(-x_{+}\right)^{m}=0.
\label{eq-serie2}
\end{equation}
Hence, we need to obtain an analytical expression for the sum of this
series in order to find exact frequency values. Such a task is very
difficult to be carried out for a general $x_{+}$, but it becomes
feasible for the large black hole limit of purely damped modes.

The coefficients $a_{m}(\omega)$ appearing in Eq. \eqref{eq-serie2}
are the Fr\"obenius coefficients of the power series solution to
either Eq. \eqref{fund-x} written for the axial functions, or the
corresponding equation in the Schwarzschild-AdS case (see
Ref. \cite{car1}). They are given in terms of the frequency, and of
the other physical parameters of the problem, by the following
recurrence formulas:
\begin{gather}
a_{1}(-x_{+})^{1}=A_{1}F_{1} a_{0}(-x_{+})^{0},
\label{rec1}\\
a_{m}(-x_{+})^{m}=\frac{F_{m}}{m}\left[A_{m}a_{m-1}(-x_{+})^{m-1}-
B_{m}a_{m-2}(-x_{+})^{m-2}\right].
\label{rec2}
\end{gather}
The last equation is valid for all $m\ge 2$ and we have
shortened the equations by introducing the quantities
\begin{gather}
A_{m}=3(m^{2}-m-1)+\left[q(q+\kappa)+(2m^{2}-2m-3)
\kappa\right]x_{+}^{2},
\label{def1}\\
B_{m}=(m^{2}-2m-3)\left[1+\kappa x_{+}^{2}\right],
\label{def2}\\
F_{m}=\left[m(3+\kappa x_{+}^{2})+2i\omega x_{+}\right]^{-1},
\label{def3}
\end{gather}
where $q=k,l$ and $\kappa=0,1$ depending on whether it is considered a plane
or spherical hole, respectively, and now the foregoing expressions
hold for any $m\ge 1$.

In the large black hole regime, the wave number (or the angular
momentum) of the perturbation is negligible and the frequency scales
as $\omega\sim 1/x_{+}$. In terms of the variable $x_{+}$, it implies
that $\omega$ has a first order pole at $x_{+}=0$. Therefore,
$\omega(x_{+})$ has a Laurent series about $x_{+}=0$, which is valid
over the range $0<x_{+}<R$, for some $R$ \cite{butkov}. Such an
expansion can be conveniently written as $-2ix_{+}\omega
=\sum_{m=0}^{\infty}b_{m}x_{+}^{m},$ where $b_{m}$ are constants to be
determined. However, not all of the elements $b_{m}$ are independent,
since part of them can be eliminated on the basis of the parity property
of the function $\omega(x_{+})$. Independently of what results from
the sum on the left-hand side of Eq. \eqref{eq-serie2}, the form of
the quantities $A_{m}$, $B_{m}$, and $F_{m}$ ensures that the product
$x_+ \omega $ is a function of $x_{+}^{2}$. This means that $\omega$
changes its sign under the transformation $x_{+}\rightarrow -x_{+}$,
being an odd function of $x_+$. Accordingly, the series representing
$x_+\omega$ should be an even function of $x_{+}$, which means that
only the even powers of $x_+$ should survive, and then one should
have $b_{2n+1}=0$ for all $n\geq 0$. This result allows us to write
\begin{equation}
-2ix_{+}\omega =
\sum_{m=0}^{\infty} c_{m}z^{m},
\label{exp2}
\end{equation}
where $z=x_{+}^{2}$ and $c_m$ are constants to
be determined.

The series representation of $2i x_{+}\omega$ opens the possibility of
Taylor expanding the terms $a_{m}(-x_{+})^{m}$ about $z=0$. Before
doing this, we notice that the value of $a_{0}$ is arbitrary, since
all the terms in the series of Eq. \eqref{eq-serie2} are proportional
to this element and the full sum is equal to zero. For simplicity, we
choose here $a_{0}=1$. Then we substitute Eq. \eqref{exp2} into
Eq. \eqref{def3} and expand $F_{m}$ (which depends on $x_+$ and also
on $\omega$) about $z=0$. When this expansion is substituted back into
Eqs. \eqref{rec1} and \eqref{rec2}, we obtain a set of Taylor series
for the elements $a_{m}(-x_{+})^{m}$ (one of such series for each
$m$), which may be represented by
\begin{equation*}
a_{m}(-x_{+})^{m}=\sum_{j=0}^{\infty}\,\eta_{m}^{j}\, z^{j},
\end{equation*}
where the elements $\eta_{m}^{j}$ are functions of $q$, $\kappa$,
$c_{0}$, $c_{1}$, ..., $c_{j}$. Substituting the previous
expansions into Eq. \eqref{eq-serie2} and collecting terms of
the same power in $z$, we find
\begin{equation}
\left[1+\eta_{1}^{0}(c_{0})+\eta_{2}^{0}(c_{0})+...\right]+
\left[\eta_{1}^{1}(c_{0},c_{1})+\eta_{2}^{1}(c_{0},c_{1})
+...\right]z+...=0.
\label{colecao}
\end{equation}

Yet the left-hand side of Eq. \eqref{colecao} is a power series in the
variable $z$. It implies that, in order for the equation being
satisfied, all bracketed terms should vanish. Therefore, the zeroth
order coefficient provides an equation for $c_{0}$, the first order
coefficient, together with the earlier calculated value of $c_{0}$,
provides an equation for $c_{1}$, and so on. Here we begin with the computations
that lead to $c_{0}$. The first two relevant terms for this task are
$\eta_{1}^{0}=-{3}/{(3-c_{0})}$ and
$\eta_{2}^{0}=-{3c_{0}}/{[2(3-c_{0})(6-c_{0})]}$. Moreover, it can be
seen from Eq. \eqref{def2} that $B_{3}$ is identically zero, and as a
consequence the term $a_{3}\,(-x_{+})^{3}$ differs from
$a_{2}\,(-x_{+})^{2}$ only by a multiplicative factor. This implies,
in particular, that the element $\eta_{3}^{0}$ is related to
$\eta_{2}^{0}$ by $\eta_{3}^{0}={5}/{(9-c_{0})}\eta_{2}^{0}.$ Such a
relation, together with the formulas \eqref{rec1} and \eqref{rec2},
implies that all the subsequent elements are also proportional to
$\eta_{2}^{0}$. More specifically, the $m$th coefficient (for $m\ge
2$) can be written as $\eta_{m}^{0}=\gamma_{m}(c_{0})\eta_{2}^{0}$,
where $\gamma_{m}(c_{0})$ is given by the recurrence formula
\begin{equation*}
\gamma_{m}(c_0)=\frac{1}{m(3m-c_{0})}\left[3(m^2-m-1)
\gamma_{m-1}(c_{0})- (m^2-2m-3)\gamma_{m-2}(c_{0})\right],
\end{equation*}
with $\gamma_{2}(c_{0})=1$ and $\gamma_{3}(c_{0})=5/9$.

Taking into account the last results, the $z$-independent term
of Eq. \eqref{colecao} gives rise to the following
equation for $c_{0}$:
\begin{equation}
c_{0}\left[\frac{1}{(3-c_{0})}+\frac{3}{2(3-c_{0})(6-c_{0})}
\sum_{m=2}^{\infty}\gamma_{m}(c_{0})\right]=0.
\label{puredampedzeros}
\end{equation}

Clearly the above equation has the special solution $c_{0}=0$, in
addition to the roots that arise as zeros of the square-bracketed
term. The existence of this special solution explains why the purely
damped modes do not scale with the horizon size, as it
happens to ordinary modes in the large black hole limit. In fact,
the numerical computations show that purely damped modes
are the only QNMs which in first approximation are not proportional to
$r_{+}$. This fact suggests they should be associated to the root
$c_{0}=0$, and then it follows from Eq. \eqref{exp2} that for large
horizon radii the corresponding frequencies go as $1/r_{+}$. On the
other hand, the numerical results also indicate that the ordinary
axial QNMs are proportional to $r_{+}$ (in first approximation). From
this one concludes that the remaining roots that are obtained from the
zeros of the term among brackets in Eq. \eqref{puredampedzeros}
correspond to the asymptotic values of the function $2ix_+\,\omega$
for these ordinary QNMs.

It is seen from the above analysis that, in order to calculate the
leading term of the large horizon limit for the purely damped mode
frequency, we must take $c_{0}=0$ and calculate the next parameter
$c_{1}$. In such a case, the coefficients $\eta_{1}^{1}$ and
$\eta_{2}^{1}$ reduce to $\eta_{1}^{1}=[q(q+\kappa)-2\kappa-c_{1}]/3$
and $\eta_{2}^{1}=-c_{1}/12$, while the other elements $\eta^1_m$ (for
$m\geq 3$) assume the form
$\eta_{m}^{1}=\gamma_{m}(0)\eta_{2}^{1}$. Substituting these results
into the first order coefficient of \eqref{colecao} we obtain an
equation that can be solved for $c_{1}$, yielding
\begin{equation*}
c_{1}=\frac{4\left[q(q+\kappa)-2\kappa\right]}{\left(4+\gamma
\right)},
\end{equation*}
where we have defined 
\begin{equation}
\gamma=\sum_{m=0}^{\infty}\gamma_m(0).
\label{gamma}
\end{equation}

The problem now is summing the series that leads to $\gamma$.
Fortunately, we succeeded in summing up this series (see Appendix B)
and the exact result is $\gamma=2$. With this, the coefficient $c_{1}$ becomes
\begin{equation*}
c_{1}=\frac{2}{3}(q-q_{-})(q-q_{+}),
\end{equation*}
where $q_{-}$ and $q_{+}$ are given by
\begin{equation}
q_{\pm}=\frac{-\kappa\pm\sqrt{\kappa^{2}+8\kappa}}{2}.
\label{parametros}
\end{equation}

Once we have obtained an expression for $c_{1}$, the
asymptotic formula of the pure imaginary QN frequency
follows directly from Eq. \eqref{exp2}. In fact,
considering a plane-symmetric AdS black hole,
for which $q=k$ and $\kappa=0$, we have $q_{-}=q_{+}=0$, and
the interesting result follows:
\begin{equation*}
\omega=i\frac{k^2}{3r_{+}}.
\end{equation*}
Alternatively, for a Schwarzschild-AdS black hole one has
$q=l$ and $\kappa=1$, which implies in $q_{-}=1$ and $q_{+}=-2$,
leading to
\begin{equation*}
\omega=i\frac{(l-1)(l+2)}{3r_{+}}\, .
\end{equation*}
Both of the last two formulas for the purely damped frequencies
are exact results at the asymptotic limit of large black holes,
in perfect agreement with the approximate
fits found through the numerical calculations.

\section{Algebraically special modes}
\label{algebrico}

The numerical results of the QN frequencies revealed a close relation
between the algebraically special modes and the purely damped
QNMs. For a fixed wave number, the imaginary frequency
$\omega=i\omega_{s}$ approaches the values of the special frequency
$\omega_{a}$ as the horizon radius goes to zero (see the comments at
the end of section \ref{perturb-NP}). There also seems to exist a
meeting point $\mbox{P}$ between the two frequencies for a special
combination of parameters such that $V^{
(-)}(r_{+})/f(r_{+})=0$, as it can be seen from the numerical results
shown in Fig. \ref{fpuro}. Furthermore, the relation between the
purely damped modes and the algebraically special frequency in the
small horizon limit is not restricted to the plane-symmetric black
hole case, but it also appears in axial perturbations in the
Schwarzschild-AdS spacetime. All these facts render it convenient to
analyze here the QN character of the algebraically special
perturbation modes. The frequency function of these modes
for the plane-symmetric black hole is given by
Eq. \eqref{algebraicfreq}, while the corresponding
expression in the spherical case is found in Ref. \cite{moss}.
In a general form, the frequency $\omega_{a}$ can be written as
\begin{equation*}
\omega_{a}=i\frac{q(q+\kappa)[q(q+\kappa)-2\kappa]}
{6(1+\kappa x_{+}^{2})}x_{+}^{3},
\end{equation*}
where again $q=k,l$ and $\kappa=0,1$, as it is considered a plane
or spherical black hole, respectively.

To verify if the algebraically special modes satisfy the QNM boundary
conditions we need to obtain the solution
$\phi_{a}^{ (-)}(x)$ associated with the algebraically
special frequency $\omega_{a}$. Since this function is a particular
case of the general power series solution \eqref{serie}, we may use
Eqs. \eqref{rec1} and \eqref{rec2} to determine the coefficients
$a_{m}$ for $\omega=\omega_{a}$. Once again we may take
$a_{0}=1$, so that the coefficients $a_{1}$ and $a_{2}$ are given,
respectively, by
\begin{equation*}
a_{1}=\frac{3(1+\kappa x_{+}^{2})-q(q+\kappa)x_{+}^{2}}{x_{+}\left(3+\kappa
x_{+}^{2}+2i\omega x_{+}\right)},
\end{equation*}
\begin{equation*}
a_{2}=\frac{x_{+}^{3}q(q+\kappa)[q(q+\kappa)-2\kappa]+6i\omega
(1+\kappa x_{+}^{2})}{4x_{+}(3+\kappa x_{+}^{2}+i\omega x_{+})
(3+\kappa x_{+}^{2}+2i\omega x_{+})}.
\end{equation*}
It is seen from the last equation that $a_{2}$ vanishes for the
frequency $\omega=\omega_{a}$. Moreover, since all the other elements
$a_{m}$ (for $m\geq 3$) differ from $a_{2}$ only by a multiplicative
factor, we have $a_{3}=a_{4}=...=0$. The solution for
$\phi_{a}^{ (-)}(x)$ then reduces to
\begin{equation}
\phi_{a}^{ (-)}(x)=1-\frac{3(1+\kappa x_{+}^{2})}{3+
\left[q(q+\kappa)+ \kappa\right]x_{+}^{2}}\left(1-\frac{x}{x_{+}}\right).
\label{afunc}
\end{equation}

We know from the Horowitz-Hubeny method that the function
\eqref{afunc} represents an ingoing wave at the event horizon. Here
the problem is verifying if $\phi_{a}^{ (-)}(x)$
satisfies the Dirichlet boundary condition at the spatial infinity.
Imposing the condition $\phi_{a}^{ (-)}(0)=0$, we
find an equation that is satisfied only if $q=q_{+}$ or $q=q_{-}$,
where $q_{+}$ and $q_{-}$ are given by Eq. \eqref{parametros}. For a
plane-symmetric AdS black hole, this means we need a vanishing
wave number in order to have a QN frequency equal to $\omega_{a}$. In
this case, however, the resulting perturbation functions would be
independent of time, which could not represent gravitational
waves. Similarly, for a Schwarzschild-AdS spacetime we need
to have $l=-2$ or $l=1$. The first root is a physically forbidden value,
while the second one leads to a static perturbation,
which only adds a small rotation to the black hole \cite{zerilli}. On
the basis of these results, we conclude that $\omega_{a}$ cannot be a QN
frequency, because the associated wave function
$\phi_{a}^{ (-)}(x)$ does not satisfy the Dirichlet
boundary condition at infinity.

Although the numerical calculations show that in the small horizon limit
$\omega=i\omega_{s}$ is very close to $\omega_{a}$, the above result
proves that the frequency function associated with purely damped modes is not
exactly equal to the algebraically special frequency. The solid line
in Fig. \ref{fpuro} does not coincide with the dots, which represent
QNMs, not even in the small black hole regime. There is, however, a
special point $\mbox{P}$ where $\omega_{a}$ seems to coincide with
$i\omega_s$. The answer to the question whether $\omega_{a}$
corresponds to a QN frequency at that point or not is
given in the following.

In spite of the generality of the analysis performed above, it fails
to test the QN character of $\omega_{a}$ in the case where
$V^{ (-)}(r_{+})/f(r_{+})=0$. This situation happens
for a special family of parameters satisfying the condition
\begin{equation}
q(q+\kappa)=\frac{3(1+\kappa x_{+}^{2})}{x_{+}^{2}}.
\label{combinacao}
\end{equation}
In this case the roots of the Fr\"obenius index equation, which are
$\alpha=0$ and $\alpha=2i\omega/f'(r_{+})=1$\, ($f' \equiv df/dr$),
differ by an integer number and the Horowitz-Hubeny method is not valid.
We can circumvent this problem by writing explicitly the axial version
of Eq. \eqref{fund-x} for $\omega=\omega_{a}$ and parameters satisfying
Eq. \eqref{combinacao}. By doing this in such a way that the analysis
holds for both the plane and the spherical AdS black holes, it is
found that
\begin{equation}
\left[(1+\kappa x_{+}^{2})x^{2}+x_{+}\,x+x_{+}^{2}\right]
\frac{d^{\,2}\phi_{a}^{ (-)}}{dx^{2}}+
\left[3(1+\kappa x_{+}^{2})x
+(3+\kappa x_{+}^{2})x_{+}\right]
\frac{d\phi_{a}^{ (-)}}{dx}-3(1+\kappa x_{+}^{2})
\phi_{a}^{ (-)}=0.
\label{fund-new}
\end{equation}
Here $x=x_+$ is a regular point of the above differential
equation, and the solution for $\phi_{a}^{
(-)}$ can be expressed as a Taylor series,
\begin{equation*}
\phi_{a}^{ (-)}(x)=
\sum_{m=0}^{\infty}d_{m}(x-x_{+})^{m}.
\end{equation*}
Substituting this expression into Eq. \eqref{fund-new}, we find as
usual that all coefficients $d_m$ (for $m\geq 2$) can be expressed in
terms of the first two, $d_0$ and $d_1$. The general solution of
Eq. \eqref{fund-new} then assumes the form
\begin{equation}
\begin{split}
&\phi_{a}^{ (-)}(x)=d_{0}\left[1+
\frac{3(1+\kappa x_{+}^{2})}{2(3+\kappa x_{+}^{2})}
\sum_{m=2}^{\infty}\frac{\beta_{m}}{(-x_{+})^{m}}
(x-x_{+})^{m}\right]\\
& +d_{1}\left[(x-x_{+})+
\frac{(3+2\kappa x_{+}^{2})}{(3+\kappa x_{+}^{2})}
\sum_{m=2}^{\infty}\frac{\beta_{m}}{(-x_{+})^{m-1}}
(x-x_{+})^{m}\right],
\label{sol-geral}
\end{split}
\end{equation}
where the parameters $\beta_{m}$ (for all $m\geq 4$) are given
by the recurrence formula
\begin{equation*}
\beta_{m}=\frac{(3+2\kappa x_{+}^{2})}{(3+\kappa x_{+}^{2})}
\beta_{m-1}-\frac{(m-3)(m+1)(1+\kappa x_{+}^{2})}{m(m-1)
(3+\kappa x_{+}^{2})}\beta_{m-2},
\end{equation*}
with $\beta_{2}=1$ and $\beta_{3}=(3+2\kappa x_{+}^{2})/(3+\kappa
x_{+}^{2})$.

The first term among brackets in Eq. \eqref{sol-geral} represents a
perturbation function which is constant at the event horizon and
corresponds to an ingoing wave, while the second term represents an
outgoing wave, since it goes as $(x-x_{+})$ in the limit $x\rightarrow
x_{+}$. To satisfy the QNM boundary condition at the horizon, we must
take $d_{1}=0$ and keep only the first term of the general solution
\eqref{sol-geral}. In addition, without loss of generality we may
choose $d_{0}=1$. Finally, we must verify if the function
$\phi_{a}^{ (-)}(x)$ satisfies the Dirichlet
boundary condition at the spatial infinity $x=0$. By substituting
$d_0=1$, $d_1=0$, and $x=0$ into Eq. \eqref{sol-geral} it follows
that
\begin{equation*}
\phi_{a}^{ (-)}(0)=1+
\frac{3(1+\kappa x_{+}^{2})}{2(3+\kappa x_{+}^{2})}
\sum_{m=2}^{\infty}\beta_{m}\equiv \phi_\infty(\kappa,x_{+})\,.
\end{equation*}

\begin{figure}
\centering\epsfig{file=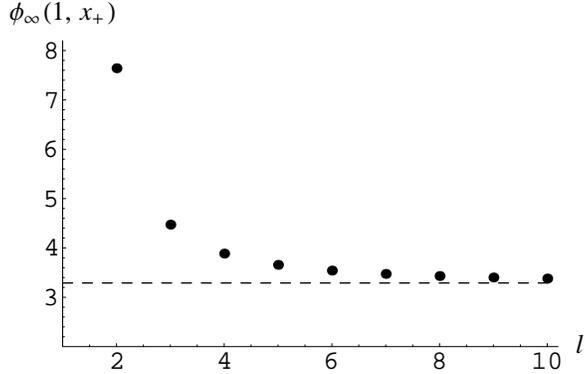}
\caption{The values of $\phi_\infty(1,x_{+})$ as a function
of the angular momentum $l$. The dots are numerical results,
while the dashed line is $\phi_\infty(0,x_{+})\simeq 3.29$.}
\label{falgeb}
\end{figure}

For plane-symmetric AdS black holes, our numerical calculations show
that $\phi_\infty (0,x_{+})\simeq 3.29$, independently of the horizon
radius. On the other hand, we obtain different values for
$\phi_\infty(1,x_{+})$ as we change the Schwarzschild black hole
size. Another way to see this variation is in terms of the angular
momentum, since $x_{+}$ and $l$ are coupled through
Eq. \eqref{combinacao}. In Fig. \ref{falgeb}, we plot the numerical
values of $\phi_\infty(0,x_+)$ (dashed line) and of
$\phi_\infty(1,x_{+})$ for $l\geq 2$ (dots). As the angular momentum
goes to infinity, which means $x_{+}\rightarrow 0$, the results found
for the spherical case tend to the value of the plane black hole.
Since $\phi_\infty(\kappa,x_{+})$ is always different from zero,
we conclude that the algebraically special modes are not QNMs,
not even for the special parameters that lead to $V^{
(-)}(r_{+})/f(r_{+})=0$. For the plane-symmetric black hole and for
$k=2$, this corresponds to the point $\mbox{P}$ in Fig. \ref{fpuro},
where the lines representing the purely damped QN and the
algebraically special frequencies tend to intercept each other. Among
other things, this result implies that the QNM spectrum of some black
holes does not present a particular pure imaginary frequency, for
those specific modes whose wave numbers (or angular momentum) satisfy
Eq. \eqref{combinacao}. For instance, an axial quadrupole ($l=2$)
perturbation is not able to excite a purely damped QNM in a
Schwarzschild-AdS black hole with $r_{+}=1$. Whether this conclusion has
some physical implication to the AdS/CFT conjecture or not is
still an open question.

\section{Final comments and conclusion}
\label{conclusao}

We have studied the gravitational quasinormal modes of plane-symmetric
anti-de Sitter black holes. These modes are responsible for the late
time behavior of small gravitational disturbances. The wave equations
governing the axial and polar perturbations were derived by a
procedure based on the Newman-Penrose formalism and on the Chandrasekhar
transformation theory. This approach allowed us to treat equally
the two spacelike coordinates $\varphi$ and $z$, as well as provided
the algebraically special frequencies associated to the gravitational
waves radiated by the black hole.

Using the Horowitz-Hubeny method, we have computed the quasinormal
frequencies for $k=2$ and several values of $r_{+}$ in the large,
intermediate, and small black hole regimes. The wave-number
dependence of the modes was obtained on the basis of the homogeneity
property of the function $\omega(r_{+},k)$. For large black holes, the
ordinary frequencies scale linearly with $r_{+}$, and the pure
imaginary frequency goes as $1/r_{+}$. For small black holes, the
fundamental modes present $\omega_{r}\rightarrow k$ and
$\omega_{i}\rightarrow 0$, while $\omega=i\omega_{s}$ approaches the
algebraically special frequency $\omega_{a}=ik^{4}/6r_{+}^{3}$.
In the limit of highly damped overtones, we found the QN frequencies
are evenly spaced, with the spacing per unit of horizon radius
being a constant which is independent of the wave-number value,
perturbation parity, and black hole size.

In relation to stability, we have shown analytically that for a
non-negative potential the imaginary part of the frequency is positive
definite for every mode. From the numerical results it also follows
that all QN frequencies have a positive imaginary part, including the
cases where the potential takes negative values. These results
conjointly imply that plane-symmetric AdS black holes are
stable against gravitational perturbations.

To study the large black hole limit we have developed an analytical
method based on the expansion of the frequency in powers of
$1/r_{+}$. Application of this method to axial perturbations showed
that there is a special frequency for which the ratio $\omega/r_{+}$
goes to zero as $r_{+}\rightarrow\infty$. The first nonvanishing term
is $\omega_{s}=ik^2/3r_{+}$ for the plane hole and
$\omega_{s}=i(l-1)(l+2)/3r_{+}$ for the Schwarzschild hole. It is
important to observe that these purely damped modes are particularly
long living for large black holes, which is exactly the relevant
regime for the AdS/CFT correspondence. If the above asymptotic
formulas could be derived on the CFT side, it could provide an
important quantitative test to the Maldacena's conjecture.

As a last remark, it should be noted that the discussion of the
physical interpretation of the modes with different wave-number
values remains. Preliminary results show that the zero wave-number axial
perturbations yield only small rotations on the system, while the
polar ones lead to an increase in the mass and the production of
gravitational waves \cite{mirandaz}. This issue, however,
must be investigated in more detail and shall be considered
in our future work.

\section*{Acknowledgments}

We are indebted to V. Cardoso for suggestions and for a critical
reading of the manuscript. We also thank J. P. S. Lemos for
stimulating conversations. One of us (A. S. M.) thanks
the Conselho Nacional de Desenvolvimento Cient{\'{\i}}fico
e Tecnol\'ogico - CNPq, Brazil, for a grant.

\begin{appendix}
\section{The pure A${\rm d}$S plane-symmetric gravitational modes}

In this appendix we discuss briefly how to find the modes of gravitational
perturbations in pure plane-symmetric AdS spacetimes. This is to be compared
to the small black hole limit QNMs, as calculated numerically in Sec.
\ref{resultados}.

The relevant wave equations for axial and polar perturbations may be obtained
from Eq. \eqref{wave} by assuming $M=0$ wherever it appears.
For instance, the potentials $V^{(\pm)}$ reduce to
the same constant value
\begin{equation}
V^{{(\pm)}} = k^2\, .
\label{pureAdSpotential}
\end{equation}
The resulting equations are then
\begin{equation}
\left(\frac{d^2}{{dr_\ast}^2} +\omega^2 - k^2\right)
 Z^{{(\pm)}}=0 \, ,
\label{pureAdSwave}
\end{equation}
whose solutions in terms of the coordinate $r=1/r_{\ast}$
are of the form
\begin{equation}
Z^{(\pm)} =\left\{
\begin{aligned}
& a^{(\pm)} \,e^{i\sqrt{\omega^2- k^2}/r} +
b^{(\pm)}\, e^{-i\sqrt{\omega^2-k^2}/r} \, ,
     \qquad & \omega^2\neq k^2 \, , \\
&\frac{c^{{(\pm)}}}{r} +
d^{{(\pm)}} \, ,\qquad &\omega^2 = k^2\, ,\\
\end{aligned}
\right.
\label{pureAdSwavefunction}
\end{equation}
where $a^{(\pm)}$, $b^{(\pm)}$,
$c^{(\pm)}$, and $d^{(\pm)}$ are
integration constants.

The solution functions for $\omega^2\neq k^2$ are analytic in
the whole range of the radial coordinate which is of interest
for comparison to the small black hole QNMs, $0< r <\infty$.
Hence, the only condition $\omega$ must
satisfy is
\begin{equation}
\omega^2> k^2 \,
\end{equation}
in order to avoid the wave growing exponentially in the limit
$r\rightarrow 0$.

However, additional restrictions to these modes arise by imposing the
usual boundary conditions for QNMs in AdS spacetimes:
no wave can come out from the black hole horizon,
which is located at $r=r_+\rightarrow 0$; and no wave can propagate
to spatial infinity, $r\rightarrow \infty$. By substituting these
conditions into Eqs. \eqref{pureAdSwavefunction} it follows that
$a^{{(\pm)}} =0$ and $b^{{(\pm)}}=0$,
which means waves whose frequencies satisfy the relation $\omega^2\neq k^2$
do vanish everywhere.

We are then left with the solutions for which $\omega^2=k^2$ only.
In this case there is in fact no wave in the radial direction,
so that the solutions $Z^{{\pm}}=
c^{(\pm)}/r +d^{(\pm)}$,
for $r> 0$, represent waves traveling along the spatial directions
orthogonal to the $r$ direction. In other words, there are
no waves coming out from the horizon nor propagating to
spacelike infinity.
Considering these as the plane-symmetric pure AdS modes,
their frequencies, given by
$$\omega =k\, ,$$
agree with the result found numerically for small black holes.

\section{Summation of the series for modes of large black holes}

Here we show how the series presented in Eq. \eqref{gamma} can be
summed up exactly. Since it is an alternating series, it is important
not only to prove its convergence, but also to test it for absolute
convergence. By mathematical induction, we find that the series
\eqref{gamma} can be written as a combination of seven terms
\begin{equation}
\gamma=2+\sum_{m=1}^{\infty}\alpha_{m}^{1}
+\sum_{m=1}^{\infty}\alpha_{m}^{2}
+\sum_{m=1}^{\infty}\alpha_{m}^{3}+\sum_{m=1}^{\infty}\alpha_{m}^{4}
+\sum_{m=1}^{\infty}\alpha_{m}^{5}+\sum_{m=1}^{\infty}\alpha_{m}^{6},
\label{comb1}
\end{equation}
where the elements $\alpha_{m}^{i}$ ($i=1$, $2$, ..., $6$)
are given by
\begin{alignat}{2}
\alpha_{m}^{1} & =(-1)^{m}\frac{2}{3^{3m+2}(6m+5)}, & \quad
\alpha_{m}^{2} & =(-1)^{m+1}\frac{2}{3^{3m}(6m+1)},\nonumber\\
\alpha_{m}^{3} & =(-1)^{m}\frac{1}{3^{3m}(6m+1)(3m+1)}, & \quad
\alpha_{m}^{4} & =(-1)^{m}\frac{6m+5}{3^{3m+2}(3m+1)(2m+1)},\nonumber\\
\alpha_{m}^{5} & =(-1)^{m}\frac{6m+5}{3^{3m+2}(3m+2)(2m+1)}, & \quad
\alpha_{m}^{6} & =(-1)^{m}\frac{12m+11}{3^{3m+2}(3m+2)(6m+5)}.\nonumber
\end{alignat}
Then we are left with a sum of six series, and by proving the absolute
convergence of each one of them, the absolute convergence of the whole
series $\sum_{\,m}\gamma_{m}(0)$ is also proved. In order to do
that, we calculate the limit
\begin{equation*}
\underset{m\rightarrow\infty}{\mbox{lim}}
\frac{\left|\alpha_{m+1}^{i}
\right|}{\left|\alpha_{m}^{i}\right|}=\frac{1}{3^3}
\end{equation*}
and verify that it is smaller than unity for any $i$, and then by
applying the D'Alembert ratio test we conclude that every individual
series is absolutely convergent. This result implies that the final
sum for $\gamma$ is independent of the element ordering in the
alternating series. In particular, we can add the terms
$\alpha_{m}^{\,i}$ over all $i$, for $i=1$, $2$, ..., $6$, before
performing the sum over $m$. By doing that, we find the simple result
$\sum_{\, i}\alpha_{m}^{i}=0$ for any value of $m$, and hence the sum
over $m$ gives zero. Therefore, the sum of the series (\ref{comb1}) is
$\gamma=2$.

\end{appendix}

\end{document}